\documentclass[12pt,preprint]{aastex}
\usepackage{graphicx}
\begin{document}
\def\la{\mathrel{\hbox{\rlap{\hbox{\lower4pt\hbox{$\sim$}}}\hbox{$<$}}}}
\def\ga{\mathrel{\hbox{\rlap{\hbox{\lower4pt\hbox{$\sim$}}}\hbox{$>$}}}}
\def\lam{$\lambda$}
\def\kms{km~s$^{-1}$}
\def\vphot{$v_{phot}$}
\def\ang{~\AA}
\def\syn{{\bf Synow}}

\title {Comparative Direct Analysis of Type~Ia Supernova Spectra.
I.~SN~1994D}

\author {David Branch, E.~Baron, Nicholas Hall, Mercy Melakayil, \&
Jerod Parrent}

\affil {Department of Physics and Astronomy, University of Oklahoma,
Norman, Oklahoma 73019, USA; e-mail: branch@nhn.ou.edu}

\begin{abstract}

As the first step in a comprehensive, comparative, direct analysis of
the spectra of Type~Ia supernovae (SNe~Ia), we use the parameterized
supernova synthetic--spectrum code, {\bf Synow}, to interpret 26
spectra of the well--observed SN~1994D.  Our results are consistent
with the traditional view that the composition structure (element
abundance fractions versus ejection velocity) is radially stratified.
We find that resonance--scattering features due to permitted lines of
Ca~II, Na~I, and Fe~II persist to more than 100 days after explosion.
The fitting parameters for SN~1994D, together with those to be
determined for other SNe~Ia, will provide an internally consistent
quantification of the spectroscopic diversity among SNe~Ia, and shed
light on how the various manifestations of observational diversity are
related to their physical causes.

\end{abstract}

\keywords{supernovae: general -- supernovae: individual (SN~1994D)}

\section{INTRODUCTION}

This is the first in a series of articles on a comprehensive,
comparative, direct analysis of spectra of Type~Ia supernovae
(SNe~Ia).  We are using the parameterized resonance--scattering
synthetic--spectrum code \syn\ to fit all of the SN~Ia spectra (other
than very late--time spectra) that are available to us.  Our
motivations include the following.  First, the process of extracting
information from supernova spectra begins with line identifications,
but the identifications of some SN~Ia spectral features remain
uncertain, and conflicting identifications can be found in the
literature.  We hope to improve this situation.  Second, we will
generate an ensemble of \syn\ fitting parameters that will provide an
internally consistent quantification of the spectroscopic differences
among SNe~Ia. This should lead to an improved understanding of how the
various manifestations of SN~Ia diversity are related to their
physical causes.

This article is concerned only with SN~1994D, a well observed,
spectroscopically normal SN~Ia that will serve as a standard of
comparison for our analyses of other SN~Ia spectra.  In forthcoming
articles we will present results for normal SNe~Ia, and then for
peculiar SNe~Ia.

In this article we confine our attention to optical spectra, from the
Ca~II H\&K blend ($gf$--weighted rest wavelength \lam3945) in the blue
to the Ca~II infrared triplet (Ca~II IR3; $gf$--weighted rest
wavelength \lam8579) in the red.  The 26 spectra selected for study
(Table~1) include 15 from Patat et~al. (1996), 8 from Filippenko
(1997), and 3 from Meikle et~al. (1996).  They range from 12 days
before to 115 days after the date of maximum light in the $B$ band,
1994 March~21.  (We also will comment on a spectrum obtained 283 days
after maximum light.)  All spectra have been corrected for the 448
\kms\ redshift of the parent--galaxy, NGC~4526, but not for
interstellar reddening, which is believed to be small.

\section{CALCULATIONS}

The elements of line formation in supernova spectra have been
discussed and illustrated by Jeffery \& Branch (1990) and Branch,
Baron \& Jeffery (2003), and will not be repeated here.  The \syn\
code is based on simple assumptions: spherical symmetry; homologous
expansion ($v=r/t$); a sharp photosphere that emits a blackbody
continuous spectrum; and line formation by resonance scattering,
treated in the Sobolev approximation.  A synthetic spectrum consists
of blended P~Cygni profiles (unshifted emission component, blueshifted
absorption component) superimposed on a continuum.  \syn\ does not do
continuum transport, it does not solve rate equations, and it does not
calculate ionization ratios.  Its main function is to take line
multiple scattering into account so that it can be used in an
empirical spirit to make line identifications and estimate the
velocity at the photosphere (or pseudo--photosphere) and the velocity
interval within which each ion is detected.  These quantities provide
constraints on the composition structure of the ejected matter.

For each ion that is introduced, the optical depth of a reference line
at one velocity (ordinarily the velocity at the photosphere) is a
fitting parameter and the optical depths of the other lines of the ion
at that velocity are calculated assuming Boltzmann excitation at
temperature $T_{exc}$.  In this article, to keep the number of fitting
parameters under control, $T_{exc}$ is taken to have the same value
for all ions at a given epoch: 10,000~K at premaximum and
near--maximum epochs and 7000~K for postmaximum epochs. All line
optical depths are taken to decrease exponentially with velocity, with
e--folding velocity $v_e$ usually taken to be 1000 \kms.  At each
epoch, the important fitting parameters are the velocity at the
photosphere, $v_{phot}$, the optical depths of the ion reference
lines, and whatever minimum and maximum velocities may be imposed on
individual ions.  When a minimum velocity is imposed that exceeds the
velocity at the photosphere, the ion is said to be detached (from the
photosphere) at that minimum velocity.  The temperature of the
blackbody continuum $T_{bb}$ is chosen to fit the overall slope of the
spectrum; its value has limited physical significance because (1) the
true continuum (or quasi--continuum) is not Planckian and (2)
interstellar extinction and wavelength--dependent errors in the
observed spectrum can affect the best fitting value of $T_{bb}$.

\section{COMPARISONS}

\subsection{Near Maximum Light}

We begin with the spectrum that is nearest to maximum light, at
day~$-1$.  In Figure~1 it is compared with a synthetic spectrum that
has $v_{phot}=11,000$ \kms, $T_{bb}=13,000$~K, and includes lines of
ten ions: O~I, Na~I, Mg~II, Si~II, Si~III, S~II, Ca~II, Fe~III, Co~II,
and Ni~II.  A value of $v_e=1000$ \kms\ is used for all ions.  Ions
responsible for absorption features in the synthetic spectrum are
labelled.  Optical depths of the reference lines are in Table~1. Note
that in addition to its photospheric component, Ca~II has a
high--velocity component, detached at 19,000 \kms.  Detached
high--velocity calcium in SN~Ia spectra was first identified by Hatano
et~al. (1999a; in SN~1994D) and has been discussed by Wang
et~al. (2003; in SN~2001el), Thomas et~al. (2004; in SN~2000cx)
Gerardy et~al. (2004; in SN~2003du), and Mazzali et~al. (2004; in
SN~1999ee).  We are confident of the presence in the observed spectrum
of O~I (here slightly detached at 12,000~\kms), Mg~II, Si~II, Si~III,
S~II, Ca~II (both photospheric and high--velocity), and Fe~III.  The
presence in the observed spectrum of Na~I, Co~II, and Ni~II is not
definite, but we use their lines in the synthetic spectrum because the
Na~I D~lines improve the fit to the observed absorption near 5770\ang,
Co~II lines improve the fit to the observed absorption near 4020\ang,
and Ni~II lines lower what otherwise would be an excessively high
Ca~II H\&K emission peak.

On the whole the fit is good, by supernova standards, although there
are discrepancies.  The optical depths of the Ca~II photospheric and
high--velocity components are chosen to fit the observed Ca~II IR3
features because, being weaker than Ca~II H\&K, they are more
sensitive to optical depth; this results in the synthetic Ca~II H\&K
absorptions (both photospheric and high--velocity) being too deep.
The discrepancy around the 5770\ang\ absorption has been encountered
in previous work with \syn.  Lines of S~II may be responsible for the
observed absorptions at 4670\ang\ and 4770\ang, but the synthetic S~II
absorptions are too weak and too blueshifted.  (Lines of
high--velocity detached Fe~II, not used in this synthetic spectrum but
definitely needed at earlier epochs (\S3.2), may also contribute to
these two observed absorptions.)  The synthetic O~I feature is
superimposed on a continuum which, as we have seen in previous \syn\
work, is too high, owing to our use of a blackbody continuum.
Additional ions would need to be introduced to beat down the synthetic
spectrum at wavelengths shorter than the Ca~II H\&K absorption.

\subsection{Premaximum}

Figure~2 shows ten premaximum spectra, including the day~$-1$ spectrum
discussed in \S3.1.  Vertical dashed lines, drawn to guide the eye,
refer to Ca~II \lam3945 blueshifted by 20,000 \kms, Si~II \lam6355
blueshifted by 10,000 \kms, and Ca~II IR3 blueshifted by 20,000 \kms.
Between day~$-1$ and day~$-9$ the spectra change slowly; between
day~$-9$ and day~$-12$ the evolution is more rapid.

In Figure~3 the day~$-9$ spectrum is compared with a synthetic
spectrum that has $v_{phot}=13,000$ \kms, $T_{bb}=12,000$~K, and
includes lines of ten ions: C~II, high--velocity Fe~II detached at
20,000 \kms, and the ions used for the day~$-1$ spectrum except for
Co~II and Ni~II.  In the synthetic spectrum both Ca~II and O~I have
detached high--velocity components.  For some of the ions a value of
$v_e$ other than 1000 \kms\ is used to improve the fit, but the $v_e$
values are not well determined.  The fit is good and all discrepancies
are mild.  We are confident of the presence in the observed spectrum
of O~I (photospheric), Mg~II, Si~II, Si~III, S~II, Ca~II (both
photospheric and high--velocity), Fe~III, and high--velocity Fe~II.
Based on the spectra of SN~1994D alone, the case for the presence of
C~II lines might not be convincing, but since we are confident of
their presence in early spectra of some other SNe~Ia, (e.g.,
SN~1998aq; Branch et~al. 2003), and they do improve the fits for
SN~1994D, we believe that they are present in SN~1994D.  The
spectroscopic case for detached high--velocity O~I, based on one weak
feature, is not compelling, but the possibility of detecting it has
been discussed by Gerardy et~al. (2004) and including it does improve
the fit.  The high--velocity Fe~II lines are discussed below, where
the need for them is more clear.

In Figure~4 the day~$-12$ spectrum is compared with a synthetic
spectrum that has $v_{phot}=14,000$ \kms, $T_{bb}=11,000$~K, and
includes lines of eight ions: the same ones used for day~$-9$ except
for Mg~II and Fe~III.  At this epoch we use $v_e = 2000$ \kms\ for
most ions.  The fit is good although there are several discrepancies.
We are confident of all identifications given in Figure~4, except for
Na~I.  Because the observed spectrum does not extend to Ca~II IR3, and
the Ca~II H\&K feature is so broad and deep that the photospheric and
detached components are badly blended, the fitting parameters for the
photospheric and detached Ca~II components are very poorly constrained.

Detached high--velocity Fe~II in early SN~Ia spectra was first
suggested by Hatano et~al. (1999a), on the basis of this day~$-12$
spectrum of SN~1994D, and it has been discussed by Branch
et~al. (2003; in SN~1998aq) and Mazzali et~al. (2005; in SN~1999ee).
These lines have a strong influence on our synthetic spectrum of
Figure~4. The long--dashed line shows the effects of removing them.
Note that in the full synthetic spectrum the S~II absorptions appear
to be weak, while in the synthetic spectrum without the high--velocity
Fe~II lines they appear stronger, even though their optical depths
have not changed.  This means that their apparent weakness in the full
synthetic spectrum, and in the observed spectrum, is not because of a
lack of optical depth in the S~II lines, but because they are
partially filled in by the P--Cygni emission components of
high--velocity Fe~II lines.

Reference--line optical depths for all of our premaximum fits,
including those that are not displayed in figures, are listed in
Table~1.  A blank entry, such as for O~I at day~$-11$, is used when
the wavelength coverage is such that an ion that might be expected to
be present, based on previous or subsequent spectra, cannot be seen.
The large changes in the photospheric and detached Ca~II optical
depths from day~$-10$ to day~$-9$ are a consequence of only the H\&K
feature being covered in the former spectrum and only the IR3 feature
in the latter.

Before leaving the premaximum phase we should mention that the
day~$-10$ and day~$-11$ spectra contain a weak absorption feature near
6630\ang\ that has a full width of only 270 \kms.  The feature
apparently is real (F.~Patat, personal communication).  We are unable
to suggest a plausible identification.  It probably is not He~I
\lam6678 because there is no corresponding feature that could be
attributed to \lam 5876, which should be stronger.  An absorption
feature due to H$\alpha$, redshifted by 3000 \kms\ and therefore
implying infall, also seems unlikely.  Similar features should be
sought in spectra of other SNe~Ia.

\subsection{The Postmaximum Si~II Phase}

Figure~5 shows eight observed spectra, from day~$-1$ (repeated from
Figure~2) to day~$+12$.  We refer to the interval from day~$+2$ to
day~$+12$ as the postmaximum Si~II phase because the 6130\ang\
absorption is deep and apparently unblended, at least in its core.
The wavelength of the absorption minimum hardly changes during this
phase.  The S~II features gradually disappear; they are still visible
(barely) at day~$+11$, but not at day~$+12$.  The absorption observed
at 5770\ang\ on day~$-1$ shifts blueward to 5690\ang\ by day~$+12$ and
develops an extended blue wing.  We attribute this to strengthening of
the Na~I D~lines.  The gradual changes at wavelengths shorter than
5200\ang\ are due to the fading of Mg~II, Si~III, and Fe~III and the
development of Co~II and Fe~II (see the reference--line optical depths
in Table~1).

Figure~6 compares the day~$+12$ spectrum with a synthetic spectrum
that has $v_{phot}=10,000$ \kms, $T_{bb}=12,000$~K, and includes lines
of eight ions: O~I, Na~I, Si~II, Ca~II, Cr~II, Fe~II, Co~II, and
Ni~II.  We use $v_e = 1000$ \kms\ for most ions, although 3000 \kms\
is used for Na~I.  The fit is not bad except for the severe problem
from 6600\ang\ to 8100\ang.  Between the Ca~II H\&K and Na~I features
the synthetic spectrum is a complex blend of many lines: mainly Co~II
and Fe~II, with some contributions from Cr~II.  In the observed
spectrum the 6130\ang\ Si~II absorption is beginning to be flanked by
weak Fe~II features (see \S3.4) that are too weak in the synthetic
spectrum.

\subsection{The Si~II--to--Fe~II Transition Phase}

At this point we switch from plotting flux per unit frequency interval
to flux per unit wavelength interval, because the spectra are becoming
redder and flux per unit wavelength interval makes the spectra look
flatter and easier to inspect.  Figure~7 shows seven observed
spectra, from day~$+12$ (repeated from Figure~5) to day~$+28$.  We
refer to the interval from day~$+14$ to day~$+28$ as the
Si~II--to--Fe~II transition phase.  At day~$+14$ the core of the deep
red Si~II absorption begins to appear blended on its red side.  At
day~$+28$ the very core of the Si~II absorption is still present, but
strongly blended on both sides, due to the strengthening of Fe~II
lines.  During this phase a striking spectral evolution occurs in the
5200\ang\ to 5600\ang\ region.  As a weak absorption at 5350\ang\
strengthens, the ratio of the flanking emission peaks evolves very
strongly, from slightly higher on the left side of the absorption at
day~$+14$ to much higher on the right side at day~$+28$.  Our fits
nicely account for this evolution in terms of strengthening Fe~II and
Cr~II lines.

Figure~8 compares the day~$+28$ spectrum with a synthetic spectrum
that has $v_{phot}=9000$ \kms, $T_{bb}=7500$~K, and includes lines of
six ions: Na~I, Si~II, Ca~II, Cr~II, Fe~II, and Co~II.  At this and
other epochs from day~14 to day~115, the imposition of a maximum
velocity of 13,000 \kms\ on Cr~II, Fe~II, and Co~II improves the
fit. Note that the two flux peaks that flank the 5350\ang\ absorption,
mentioned in the previous paragraph, are well matched.  The spectrum
between the Ca~II H\&K and Na~I features remains a complex blend of
many lines, but now mainly Fe~II, with some contributions from Co~II
and Cr~II.  We believe that the line identifications in Figure~8 are
correct as far as they go, but we have not found a plausible way to
remove the discrepancy to the right of the 6500\ang\ flux peak.

\subsection{The Fe~II Phase}

Figure~9 shows five observed spectra, from day~$+28$ (repeated from
Figure~7) to day~$+283$.  We refer to the interval from day~$+50$ to
day~$+115$ as the Fe~II phase, because the Si~II line is all but gone
(in our synthetic spectrum for day~$50$, Si~II detached at 9000~\kms\
still has a small effect) and apart from the persistent Ca~II and Na~I
features the spectrum is mainly shaped by Fe~II lines.  Most of the
spectral features retain their identities and evolve slowly during
this phase, although the Na~I feature becomes much stronger, and an
apparent emission peak develops at 7370\ang.  The day~$+50$ spectrum
can be fit about as well as the day~$+28$ spectrum (Figure~8), in much
the same way.

Figure~10 compares the day~$+115$ spectrum with a synthetic spectrum
that has $v_{phot}=6000$ \kms, $T_{bb}=15,000$~K, and includes lines
of only four ions: Na~I, Ca~II, Fe~II, and Co~II.  The high value of
$T_{bb}$ was chosen on the basis of wavelengths longer than 4500\ang;
this leaves the synthetic spectrum much too high at wavelengths
shorter than 4400\ang.  The discrepancy may be due to a strong
departure from an underlying blackbody continuum, and to severe
underblanketing in the synthetic spectrum.  (Lines of neutral
iron--group elements may help to blanket the observed spectrum at this
epoch; see Pastorello et~al. (2004) on the spectrum of the Type~II
SN~1998A.)  The synthetic spectrum also fails completely from
6600\ang\ to 7600\ang.  Nevertheless, the fit from 4500\ang\ to
6500\ang, although not good, is sufficient to suggest that, contrary to
what is usually assumed, the spectrum at 100~days postmaximum is not
just a blend of forbidden emission lines --- it is largely shaped by
permitted Fe~II lines, especially in the blue.

The overall shape of the day~$+283$ spectrum (Figure~9) differs from
that of the day~$+115$ spectrum, and the assumption of an underlying
blackbody continuum at day~$+283$ would be completely untenable, yet
most of the flux minima are at practically the same wavelengths in
these two spectra.  If the day~$+283$ spectrum consisted just of
forbidden emission lines, with no absorption features, these
similarities would be strangely coincidental.  However, Ca~II H\&K
absorption is apparent in the day~$+283$ spectrum, and Ca~II IR3 and
Na~I absorptions appear to be present as well.  In addition, the sharp
absorption at 4100\ang\ may be produced by Ca~I \lam4226 (a resonance
line), blueshifted by 8900 \kms.  The extent to which SN~Ia spectra as
late as day~$+283$ are affected by Fe~II permitted lines, as well,
will be addressed in a future article, when we have the opportunity
to study spectra in the interval between day~$+115$ and day~$+283$.

\section{DISCUSSION}

We have been able to identify nearly all of the spectral features of
SN~1994D with lines of plausible ions --- mostly in the same way they
have been identified in spectra of normal SNe~Ia before (e.g., Branch
et~al. 1983; Jeffery et~al. 1992; Kirshner et~al. 1993; Mazzali
et~al. 1993; Branch et~al. 2003). Our fits to all but the latest
spectra are satisfactory, considering the approximations made, which
include spherical symmetry.  Qualitatively, our results are consistent
with the traditional view that the SN~Ia composition structure is
strongly radially stratified (Figure~11), e.g., C~II only above
13,000~\kms\ and O~I only above 12,000~\kms.

The reference--line optical depths of Table~1 could be used, together
with LTE line--optical--depth calculations such as those of Hatano
et~al. (1999b), to estimate element abundance ratios, but these would
be subject to serious uncertainties.  In general, quantitative
inferences about element abundances should be made by means of
detailed spectrum calculations for nuclear--hydrodynamical explosion
models that include complete composition structures (e.g., H\"oflich
1995 and Lentz et al. 2001 for SN~1994D).  However, a few of the
quantitative implications of our present results should be mentioned.
Since we have invoked the presence of permitted Fe~II, Ca~II, and Na~I
 features at relatively late times, we make rough estimates of
the amount of mass in these ionization states that would be needed to
produce the optical depths.  For a given ionization stage, the amount
of mass required to fill a uniform--density sphere of radius $vt$ at
time $t$ after explosion is

$$ M(ion)/M_\odot \simeq 10^{-14}\ v_4^3\ t_d^2\ A\ \tau / f\
\lambda_\mu\ x_\ell, $$

\noindent where $v_4$ is in units of $10^4$ \kms, $t_d$ is in days,
$A$ is atomic weight, $f$ is oscillator strength, $\lambda_\mu$ is in
microns, and $x_\ell$ is the fraction of the ion population in the
lower level of the transition.  With $v_4=0.6$ and $t_d=100$, the
amount of singly ionized iron required to produce a reference--line
optical depth of 50 is about 0.01~M$_\odot$.  The amount of singly
ionized calcium required to produce a reference--line optical depth of
$10^4$ is about $4 \times 10^{-4}$~M$_\odot$. (For comparison, the
total amount of calcium in the sun is $6.4 \times 10^{-5}$~M$_\odot$.)
The amount of neutral sodium required to produce a reference--line
optical depth of 10 is $3\times 10^{-7}$~M$_\odot$. (The total amount
of sodium in the sun is $3.4 \times 10^{-5}$~M$_\odot$.)  These mass
requirements are modest, provided that the fractions of singly ionized
iron, singly ionized calcium, and neutral sodium, are not extremely
small.

The relative optical depths of Ni~II, Co~II, and Fe~II, and of Ni~III,
Co~III, and Fe~III, do not depend critically on temperature and
electron density, therefore we ask whether the relative optical
depths that we have used for these ions appear to be reasonable, on
the basis of LTE line optical depth calculations.  Briefly, we find
the following: (1) The iron responsible for the detached
high--velocity features at early times was not formed from the decay
of $^{56}$Ni through $^{56}$Co, otherwise Ni~II and Co~II lines would
be too strong. (2) There is no conflict between the presence of Fe~III
lines and the absence of Co~III and Ni~III lines from day~$-4$ to
day~7. (3) The ratios of the postmaximum Co~II and Fe~II optical
depths are roughly consistent with cobalt and iron both having been
formed from $^{56}$Ni decay (or, more reasonably, with small amounts
of directly synthesized iron, cobalt, and nickel such that their final
relative abundances match their relative abundances in the sun). (4)
If the Ni~II identifications from day~10 to day~17 are correct, then
the nickel responsible for these features cannot have been formed as
$^{56}$Ni, but must have been synthesized directly, as stable
$^{58}$Ni.

The optical depths listed in Table~1 indicate which ions are
responsible for shaping the spectrum of SN~1994D at each epoch.  The
main utility of these parameters, however, will be to serve as a
standard of comparison for other SNe~Ia.

We are grateful to Nando Patat, Alex Filippenko, and Peter Meikle for
providing spectra, and to David Jeffery and Jason Zinn for assistance.
This work has been supported by NSF grant AST-0204771 and NASA LTSA
grant NNG04GD36G.

\clearpage     

\begin{figure}
\includegraphics[width=.7\textwidth,angle=270]{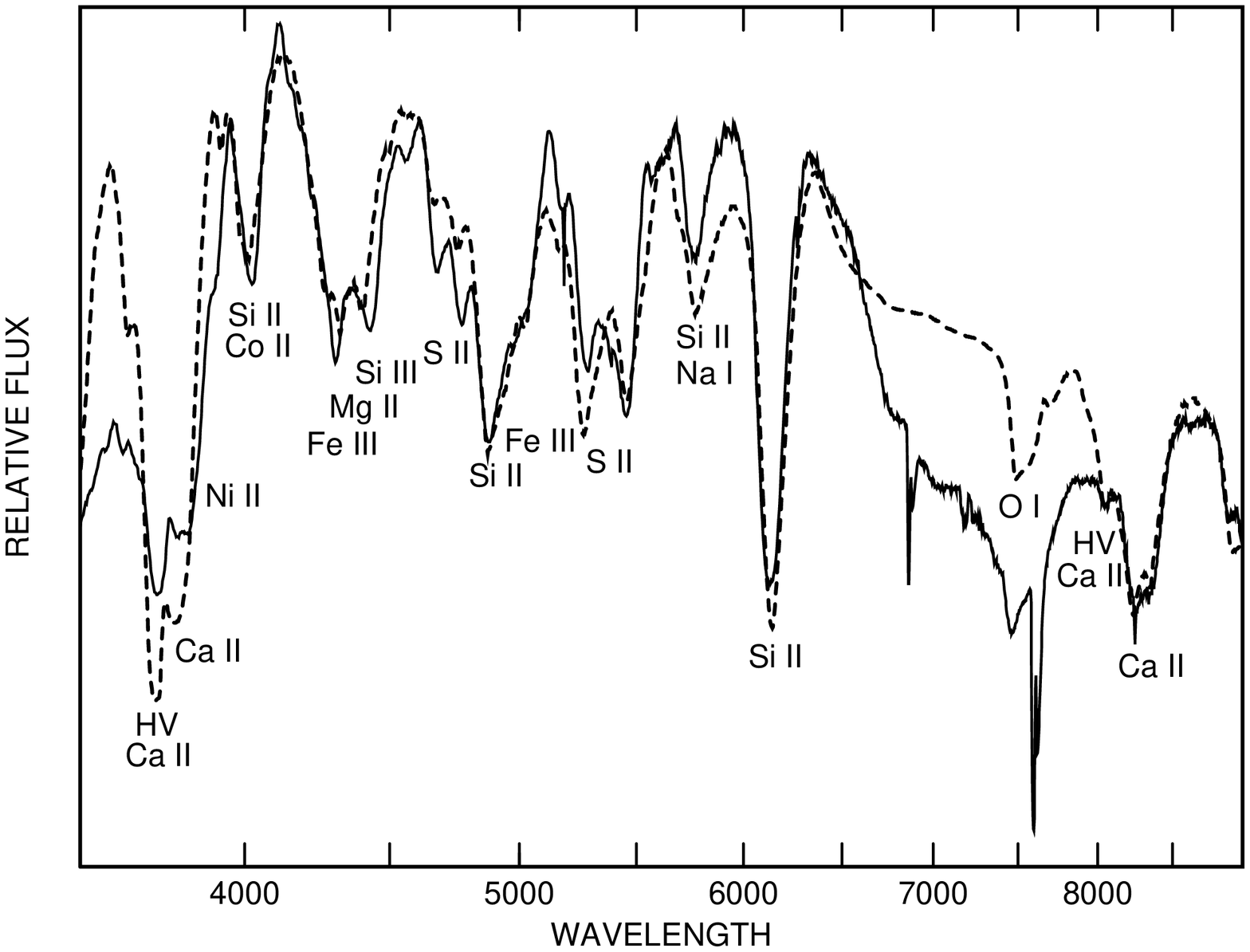}
\caption{The day~$-1$ spectrum of SN~1994D ({\sl solid line}) is
  compared with a synthetic spectrum ({\sl dashed line}) that has
  $v_{phot}=11,000$ \kms, $T_{bb}=13,000$~K, and includes lines of ten
  ions. The flux is per unit frequency interval.}
\end{figure}

\begin{figure}
\includegraphics[width=.8\textwidth,angle=0]{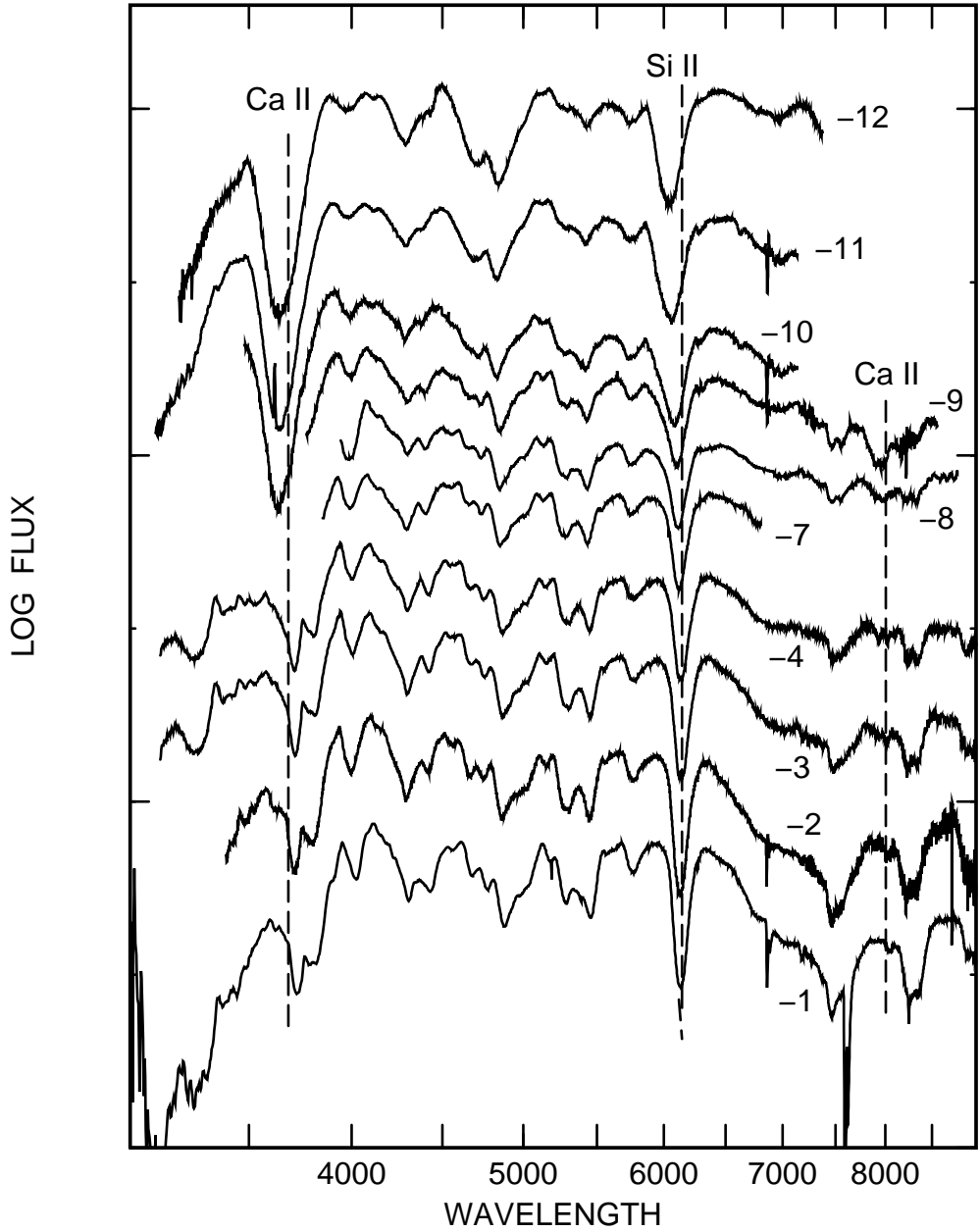}
\caption{Ten spectra of SN~1994D.  Epochs are in days
  with respect to the date of maximum brightness in the $B$ band.  The
  flux is per unit frequency interval and the vertical displacement is
  arbitrary.  Vertical dashed lines refer to Ca~II \lam3945
  blueshifted by 20,000 \kms, Si~II \lam6355 blueshifted by 10,000
  \kms, and Ca~II IR3 blueshifted by 20,000 \kms.}
\end{figure}

\begin{figure}
\includegraphics[width=.7\textwidth,angle=270]{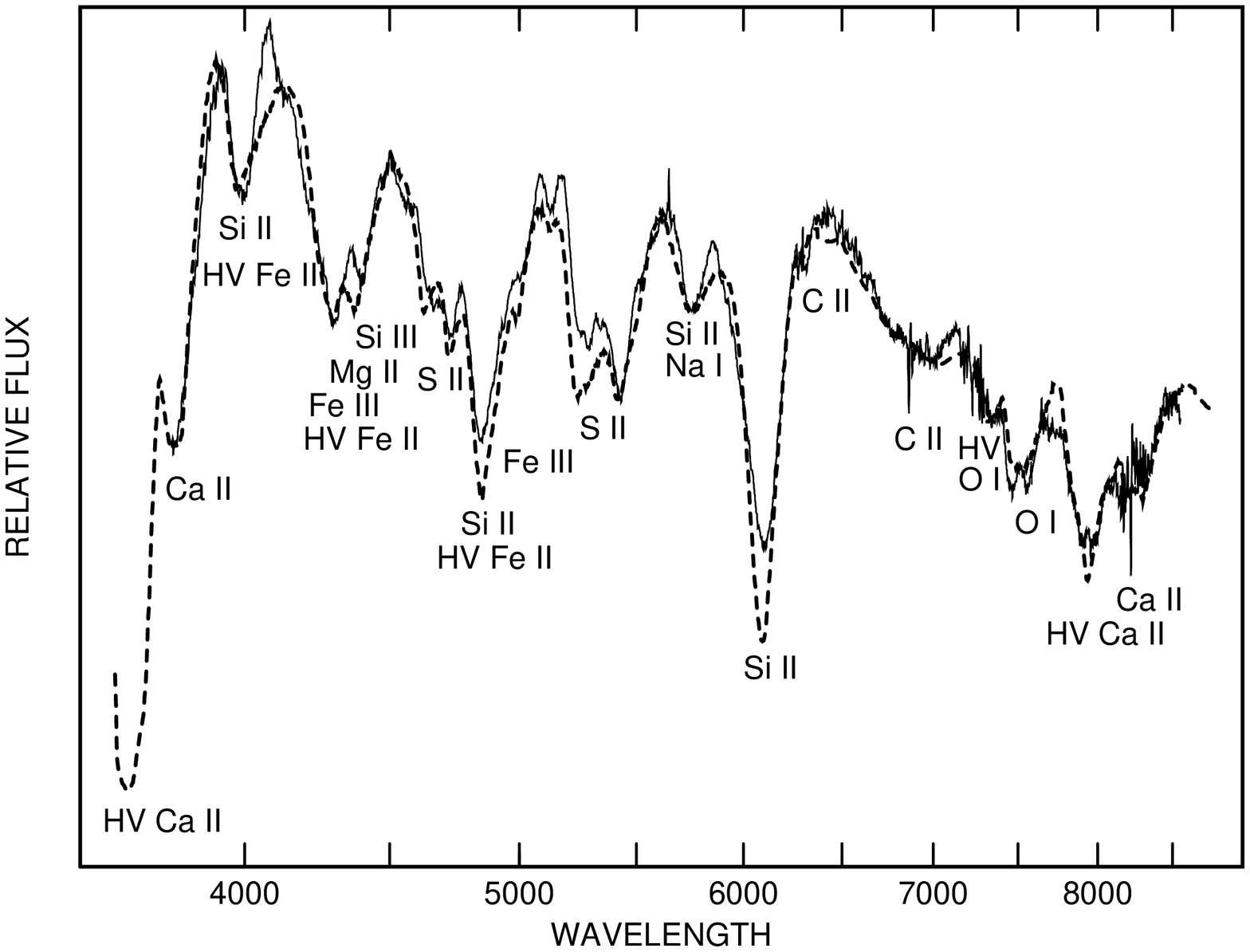}
\caption{The day~$-9$ spectrum of SN~1994D ({\sl solid line}) is
  compared with a synthetic spectrum ({\sl dashed line}) that has
  $v_{phot}=13,000$ \kms, $T_{bb}=12,000$~K, and includes lines of
  ten ions. The flux is per unit frequency interval.}
\end{figure}

\begin{figure}
\includegraphics[width=.7\textwidth,angle=270]{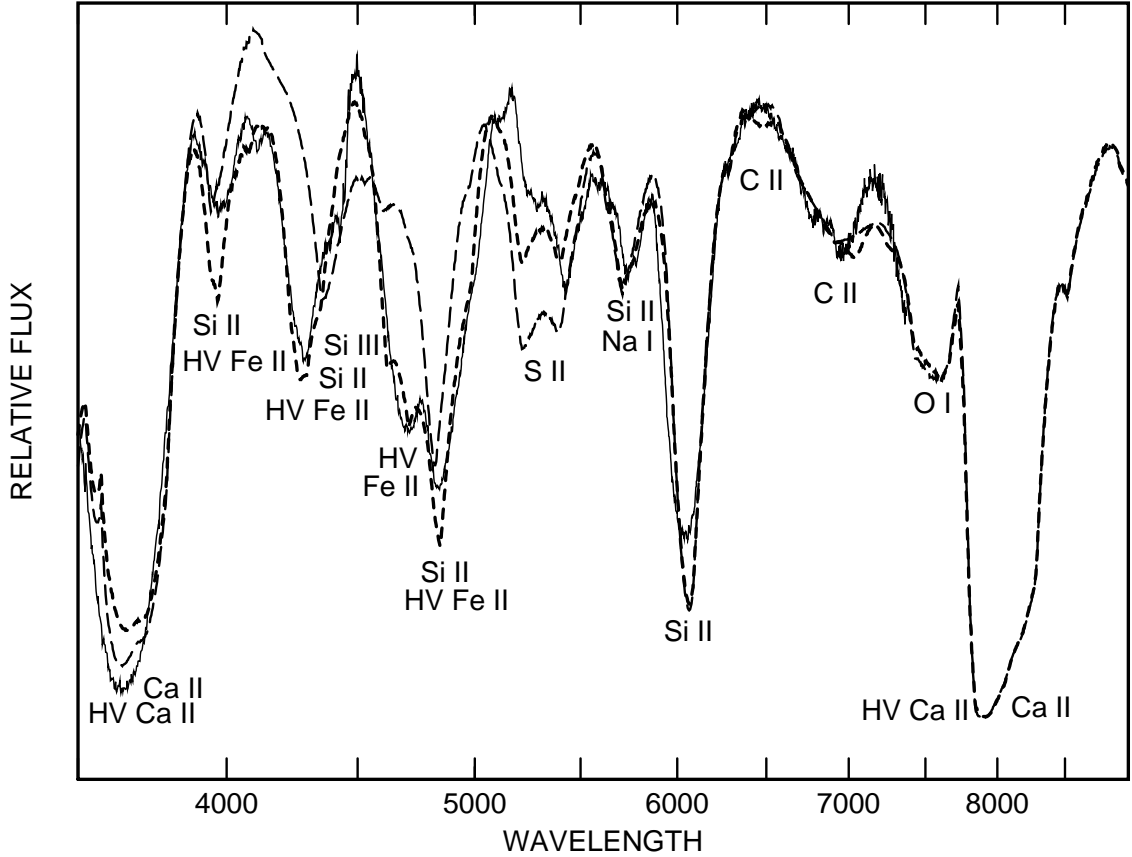}
\caption{The day~$-12$ spectrum of SN~1994D ({\sl solid line}) is
  compared with a synthetic spectrum ({\sl short--dashed line}) that
  has $v_{phot}=14,000$ \kms, $T_{bb}=11,000$~K, and includes lines of
  eight ions, and with the same synthetic spectrum without the
  high--velocity Fe~II lines ({\sl long--dashed line}). The flux is
  per unit frequency interval.}
\end{figure}

\begin{figure}
\includegraphics[width=.8\textwidth,angle=0]{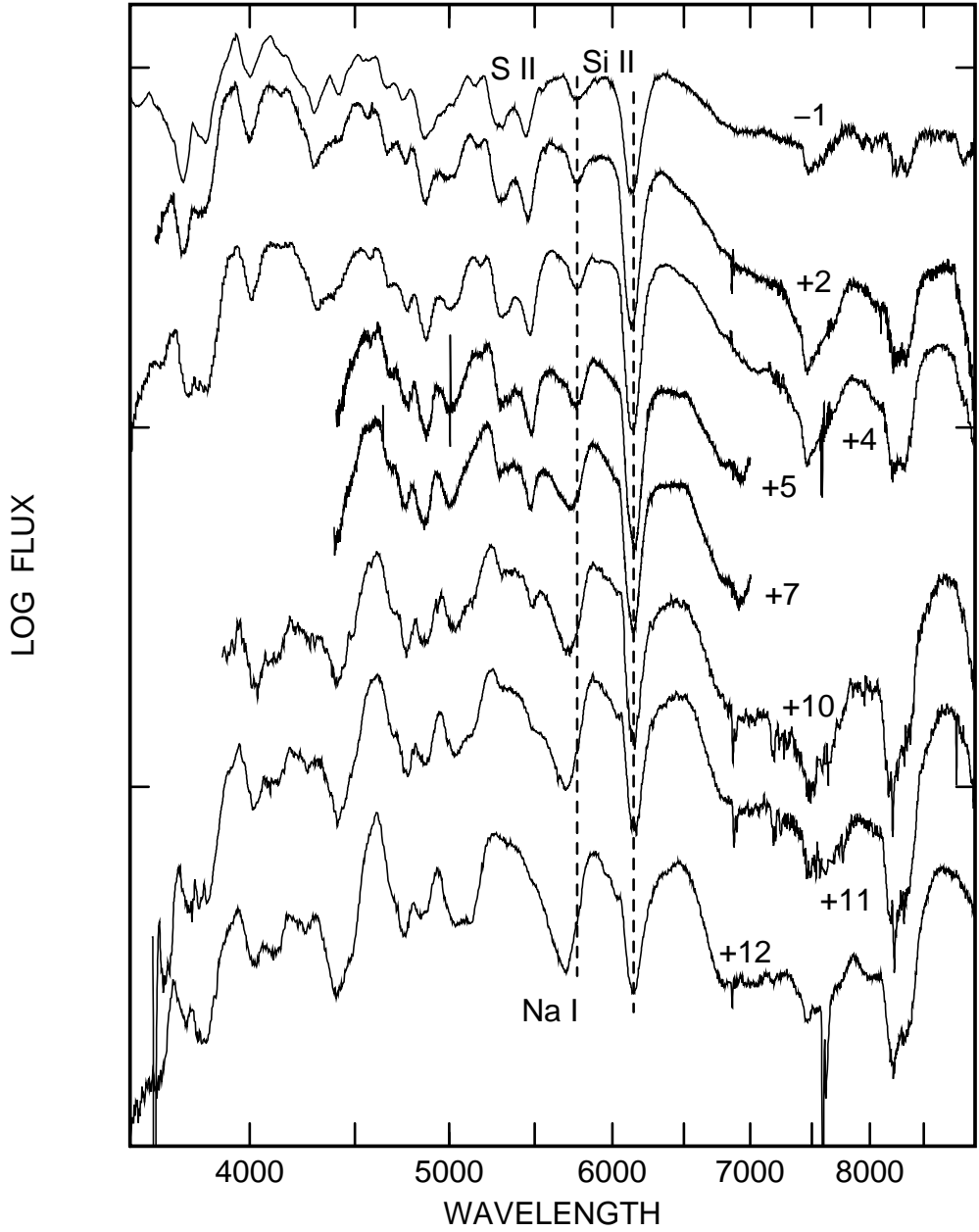}
\caption{Eight spectra of SN~1994D.  The flux is per unit frequency
  interval.  Vertical dashed lines refer to Si~II \lam5972 and
  \lam6355, both blueshifted by 10,000 \kms.}
\end{figure}

\begin{figure}
\includegraphics[width=.8\textwidth,angle=270]{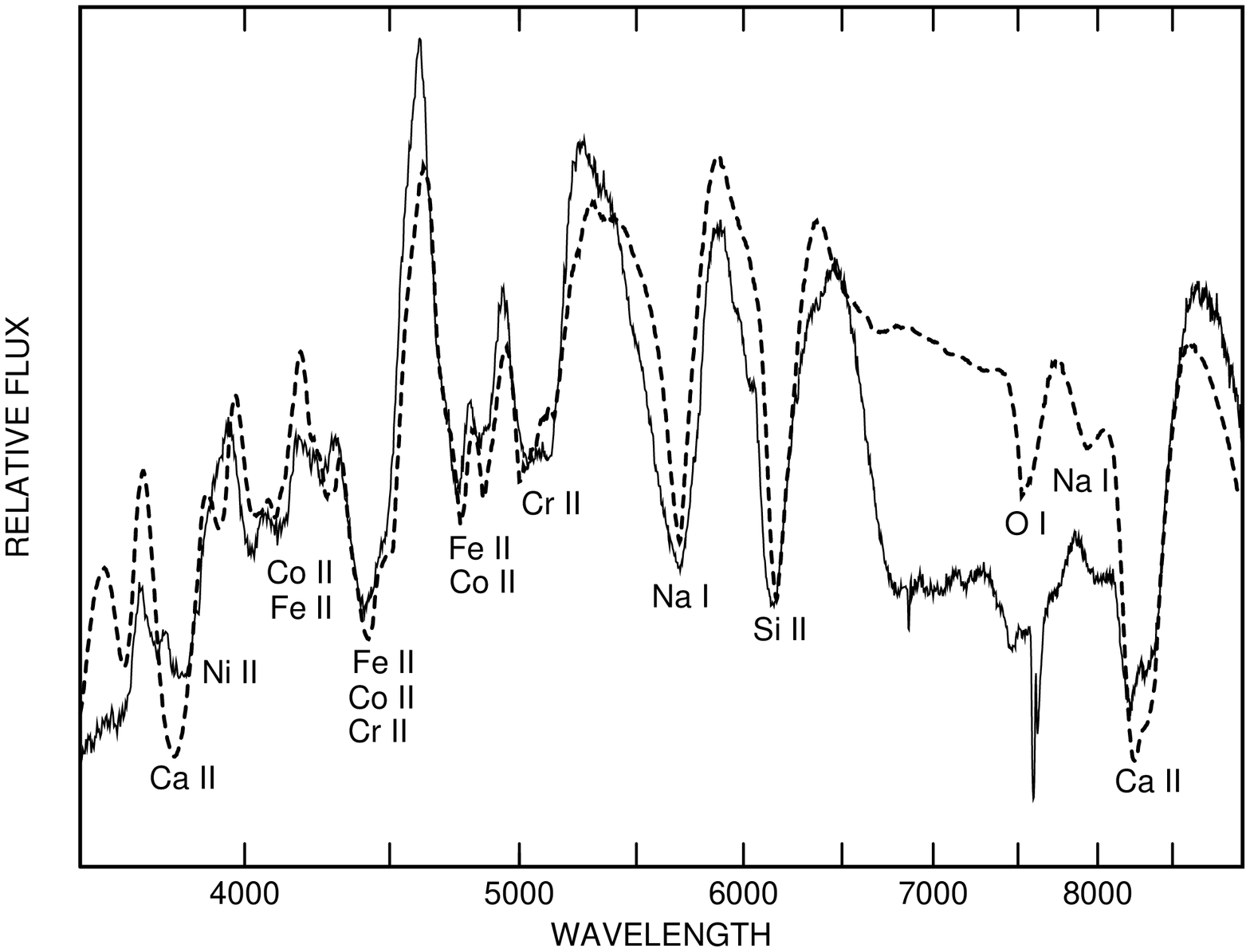}
\caption{The day~$+12$ spectrum of SN~1994D ({\sl solid line}) is
  compared with a synthetic spectrum ({\sl dashed line}) that has
  $v_{phot}=10,000$ \kms, $T_{bb}=12,000$~K, and includes lines of
  eight ions. The flux is per unit frequency interval.}
\end{figure}

\begin{figure}
\includegraphics[width=.8\textwidth,angle=0]{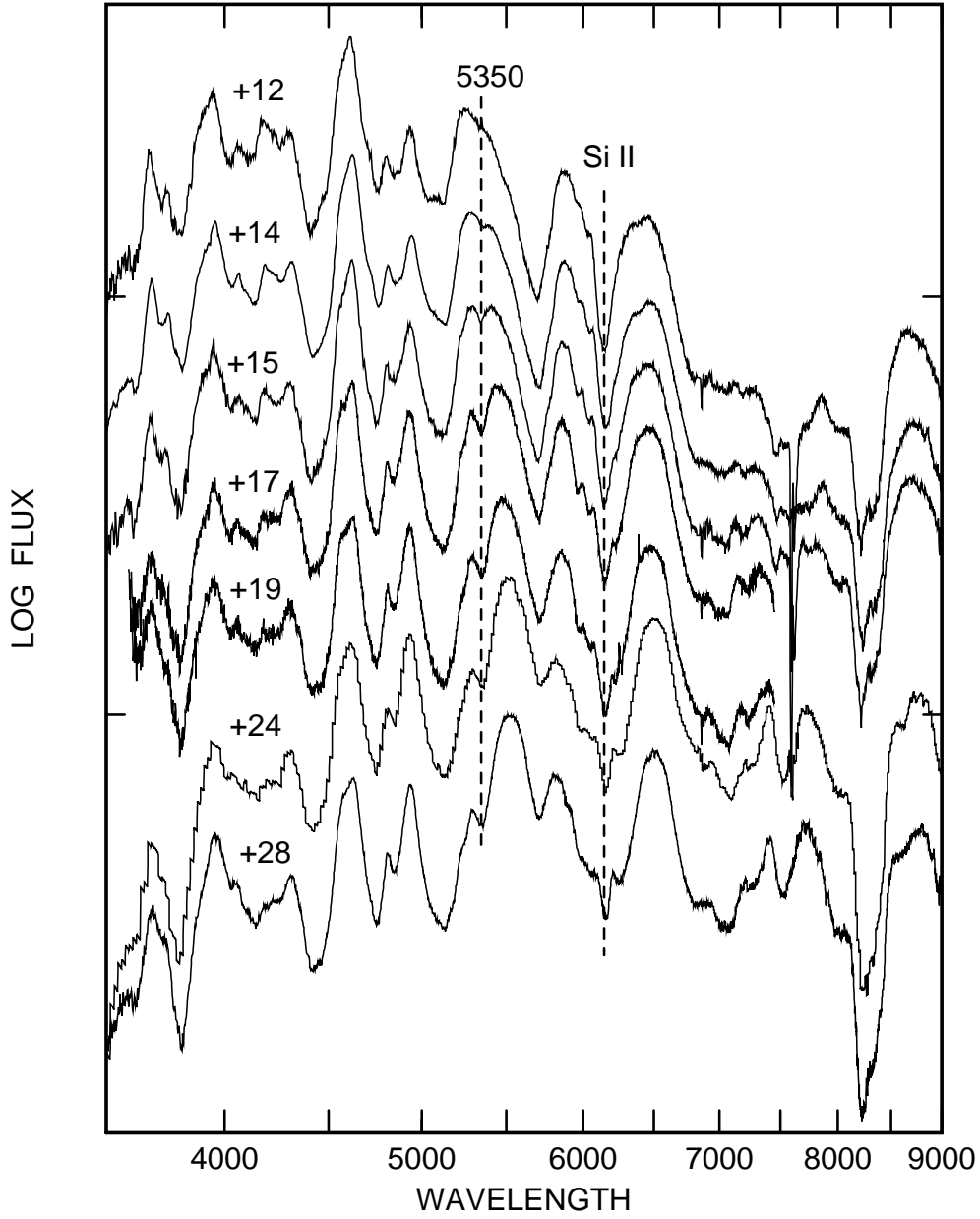}
\caption{Seven spectra of SN~1994D.  The flux is per unit wavelength
  interval. One vertical dashed line is at 5350\ang\ and the other
  refers to Si~II \lam6355 blueshifted by 10,000 \kms.}
\end{figure}

\begin{figure}
\includegraphics[width=.8\textwidth,angle=270]{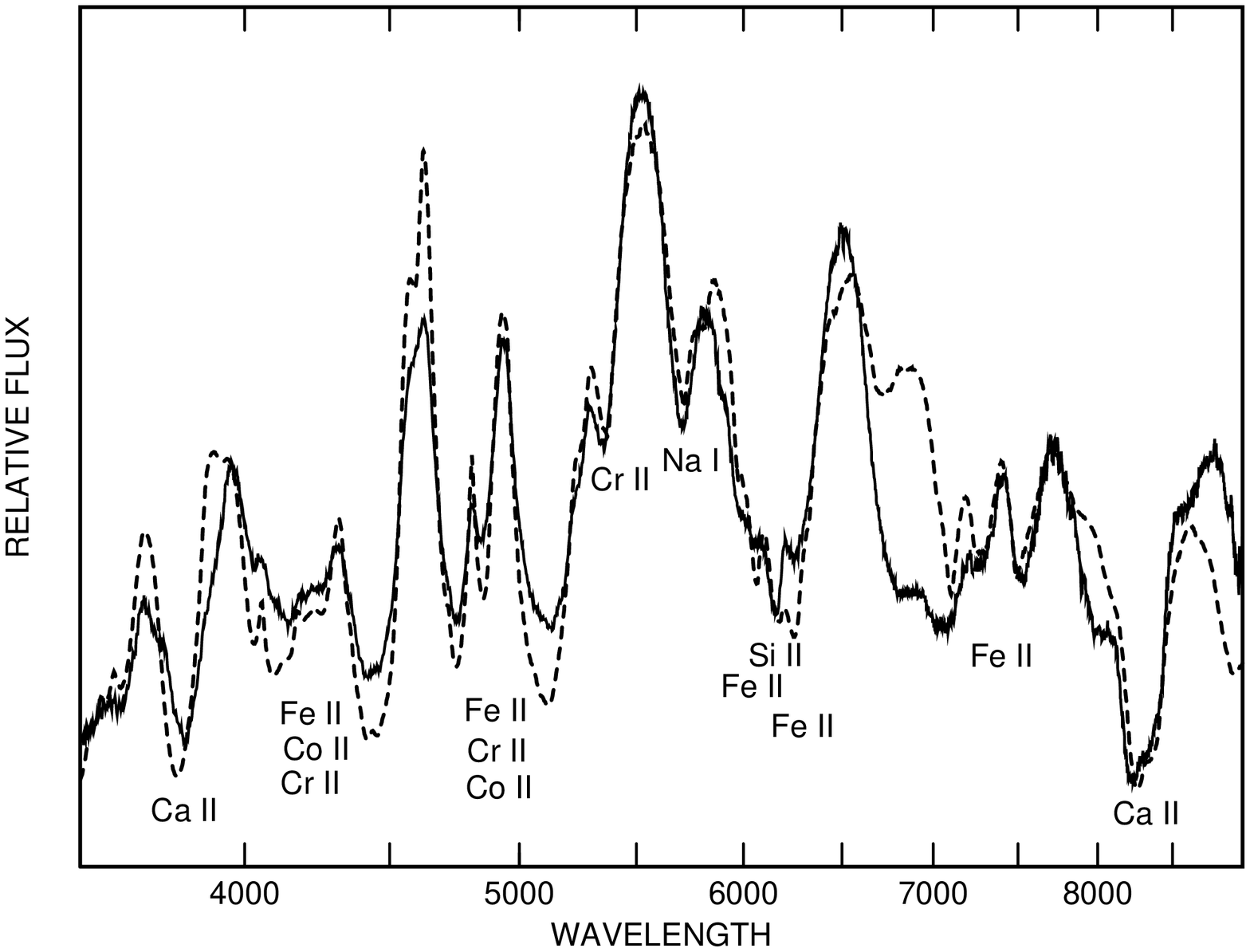}
\caption{The day~$+28$ spectrum of SN~1994D ({\sl solid line}) is
  compared with a synthetic spectrum ({\sl dashed line}) that has
  $v_{phot}=9000$ \kms, $T_{bb}=7500$~K, and contains lines of six
  ions.  The flux is per unit wavelength interval.}
\end{figure}

\begin{figure}
\includegraphics[width=.8\textwidth,angle=0]{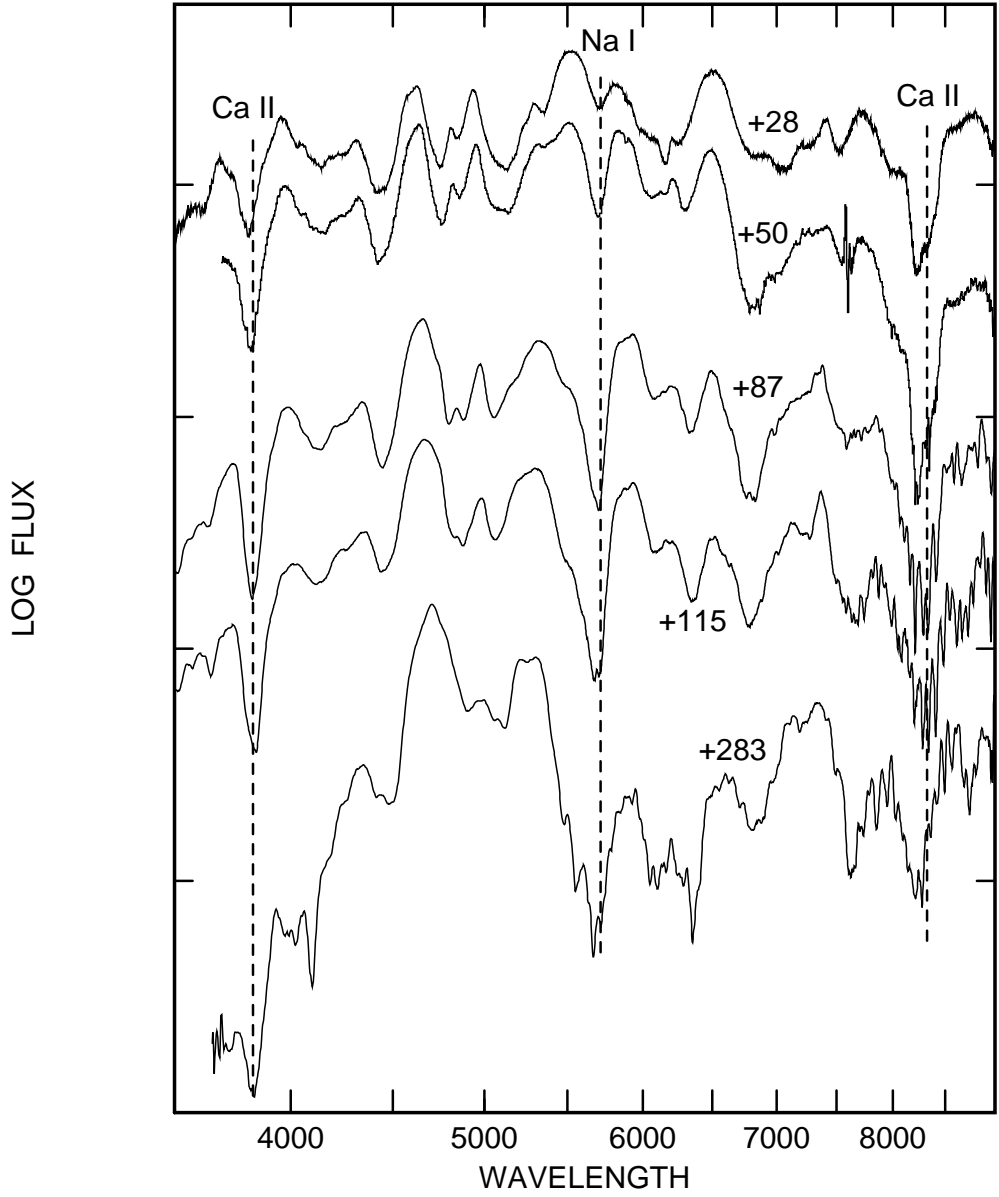}
\caption{Five spectra of SN~1994D.  The flux is per unit wavelength
  interval. Vertical dashed lines refer to Ca~II \lam3945, Na~I
  \lam5892, and Ca~II IR3, each blueshifted by 9000 \kms.}
\end{figure}

\begin{figure}
\includegraphics[width=.8\textwidth,angle=270]{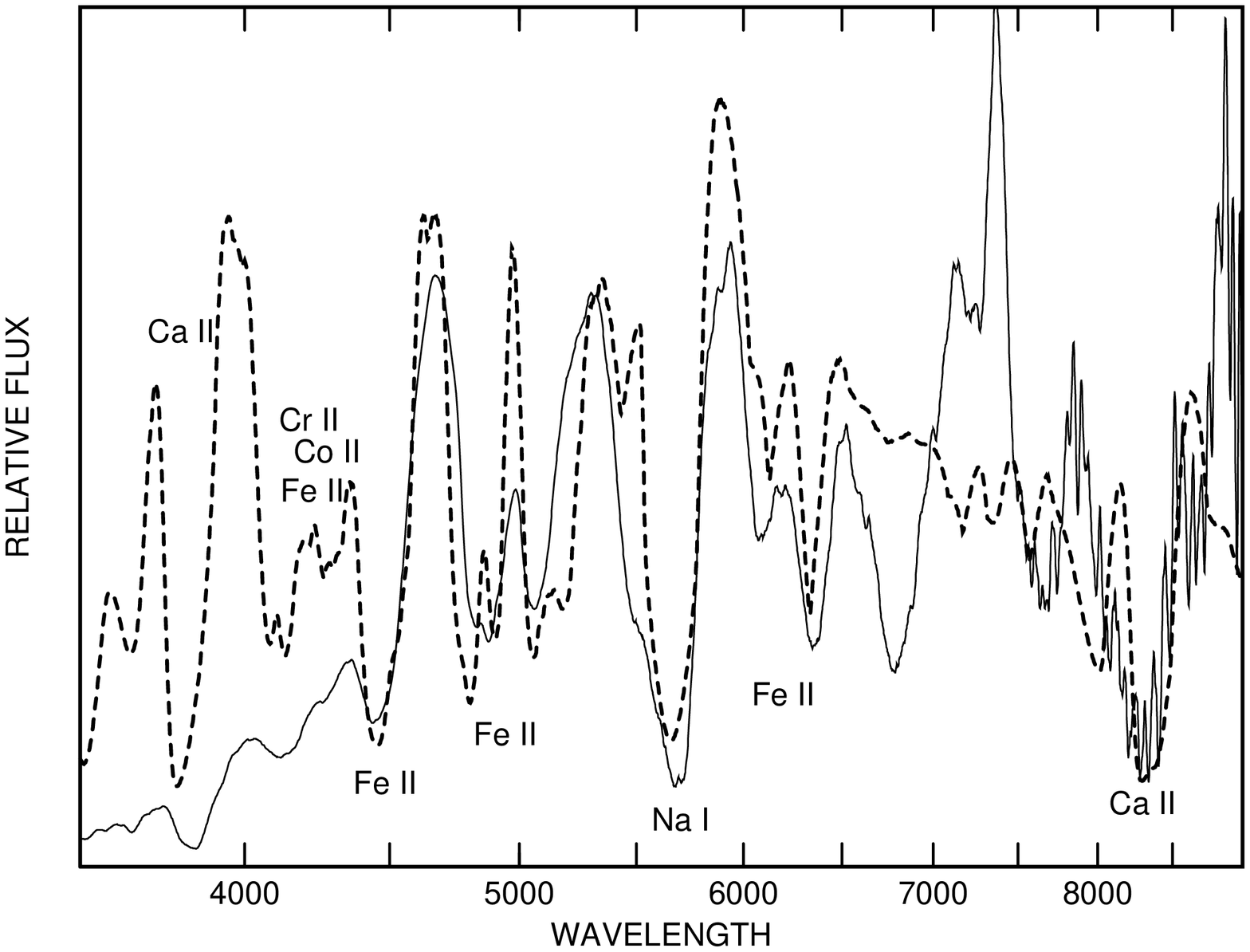}
\caption{The day~$+115$ spectrum of SN~1994D ({\sl solid line}) is
  compared with a synthetic spectrum ({\sl dashed line}) that has
  $v_{phot}=6000$ \kms, $T_{bb}=15,000$~K, and contains lines of four
  ions.  The flux is per unit wavelength interval.}
\end{figure}

\begin{figure}
\includegraphics[width=.8\textwidth,angle=0]{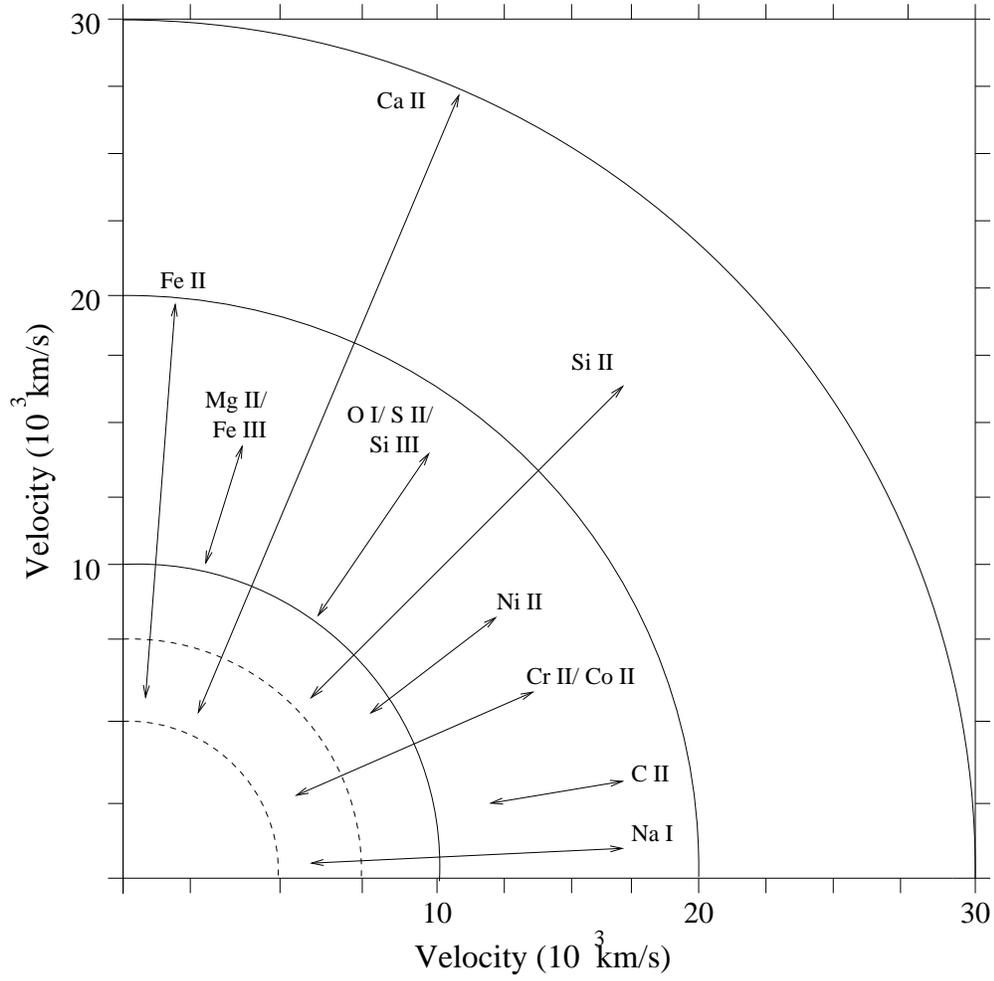}
\caption{The velocity intervals within which the ions were used in the
  synthetic spectra.  The upper bound is the velocity at which the
  line optical depth fell below 0.1.}
\end{figure}

\clearpage

\begin{deluxetable}{ccccccccccccccccccccccc}
\tabletypesize{\scriptsize}
\rotate
\tablenum{1}
\setlength{\tabcolsep}{0pt}

\tablecaption{Input Parameters for SN~1994D}

\tablehead{\colhead{day} & \colhead{$v_{phot}$} &
\colhead{$T_{bb}$} & \colhead{$v_e$} & \colhead{$T_{exc}$} &
\colhead{$\tau$} & \colhead{$\tau$} & \colhead{$\tau$} &
\colhead{$\tau$} & \colhead{$\tau$} & \colhead{$\tau$} &
\colhead{$\tau$} & \colhead{$\tau$} & \colhead{$\tau$} &
\colhead{$\tau$} & \colhead{$\tau$} & \colhead{$\tau$} &
\colhead{$\tau$} & 
\colhead{$\tau$}\\ \colhead{} & \colhead{$10^3$ km~s$^{-1}$} & \colhead{K} &
\colhead{$10^3$ km~s$^{-1}$} & \colhead{K} & \colhead{C~II} & \colhead{O~I} &
\colhead{Na~I} & \colhead{Mg~II} & \colhead{Si~II}
& \colhead{Si~III} & \colhead{S~II} & \colhead{Ca~II} &
\colhead{HV~Ca~II} & \colhead{Cr~II} & \colhead{Fe~II} &
\colhead{Fe~III} & \colhead{Co~II} &
\colhead{Ni~II} }

\startdata

-12 & 14 & 11 & 2 & 10 & 0.6 & 0.5 & 0.5 & 0 & 15 & 0.7 & 1
& 600 & 200 & 0 & 1.5 & 0 & 0 & 0\\

-11 & 14 & 11 & 2 & 10 & 0.6 &  & 0.5 & 0 & 7 & 0.5 & 1
& 600 & 200 & 0 & 0.8 & 0 & 0 & 0\\

-10 & 13 & 12 & 2 & 10 & 0.6 &  & 0.5 & 0 & 7 & 0.5 & 1
& 600 & 200 & 0 & 0.8 & 0 & 0 & 0\\

-9 & 13 & 12 & 1 & 10 & 0.5 & 1 & 0.3 & 0.4 & 5 & 1& 2
& 20 & 5 & 0 & 0.4 & 0.4 &0 & 0\\

-8 & 12 & 12 & 1 & 10 & 0.5 & 0.8 & 0.3 & 0.5 & 7 & 0.8 & 1.5
& 8 & 3 & 0 & 0 & 0.5 & 0 & 0\\

-7 & 12 & 12 & 1 & 10 & 0.4 &  & 0.3 & 1& 7 & 0.8 & 1.5
&  &  & 0 & 0 & 0.6 & 0 & 0\\

-4 & 11 & 13 & 1 & 10 & 0.2 & 0.5 & 0 & 1& 8 & 0.8 & 1.8
& 15 & 2 & 0 & 0 & 0.8 & 0.2 & 0\\

-3 & 11 & 13 & 1 & 10 & 0 & 0.5 & 0.3 & 1& 9 & 0.8 & 1.8
& 40 & 5 & 0 & 0 & 0.8 & 0.4 & 0\\

-2 & 11 & 13 & 1 & 10 & 0 & 0.8 & 0.4 & 1& 10 & 0.8 & 2
& 70 & 5 & 0 & 0 & 0.8 & 0.6 & 0.6\\

-1 & 11 & 13 & 1 & 10 & 0 & 0.8 & 0.4 & 1& 12 & 1.2 & 2
& 70 & 5 & 0 & 0 & 0.8 & 0.8 & 0.8\\

2 & 11 & 35 & 1 & 10 & 0 & 2 & 0.15 & 1.5 & 12 & 0.8 & 2.5
& 70 & 2 & 0 & 0 & 0.8 & 0.5 & 0\\

4 & 10 & 11 & 1 & 7 & 0 & 1.5 & 0.15 & 1.5 & 20 & 0.8 & 2
& 50 & 0 & 0 & 0.5 & 0.8 & 0.5 & 0\\

5 & 10 & 14 & 1 & 7 & 0 &  & 0.3 &  & 15 & 0.8 & 1.5
&  & 0 & 0 & 1& 0.8 & 1& \\

7 & 10 & 14 & 1 & 7 & 0 &  & 0.8 &  & 10 & 1& 1
&  & 0 & 0 & 2 & 1& 5 & \\

10 & 10 & 12 & 1 & 7 & 0 & 2 & 1.5 & 0 & 10 & 0 & 1
& $10^3$ & 0 & 1.5 & 4 & 0 & 10 & 8\\

11 & 10 & 12 & 1 & 7 & 0 & 1 & 2 & 0 & 10 & 0 & 0.3
& $2\times10^3$ & 0 & 1.5 & 5 & 0 & 10 & 10\\

12 & 10 & 12 & 1 & 7 & 0 & 1 & 1.5 & 0 & 8 & 0 & 0
& $2\times10^3$ & 0 & 2 & 6 & 0 & 12 & 15\\

14 & 10 & 11 & 1 & 7 & 0 & 0 & 2 & 0 & 10 & 0 & 0
& $2\times10^3$ & 0 & 5 & 20 & 0 & 30 & 25\\

15 & 10 & 9 & 1 & 7 & 0 & 0 & 1.5 & 0 & 10 & 0 & 0
& $2\times10^3$ & 0 & 5 & 50 & 0 & 30 & 15\\

17 & 10 & 8 & 1 & 7 & 0 & 0 & 1.5 & 0 & 10 & 0 & 0
& 500 & 0 & 20 & 70 & 0 & 30 & 5\\

19 & 9 & 8.5 & 1 & 7 & 0 & 0 & 1 & 0 & 6 & 0 & 0
& 500 & 0 & 40 & 100 & 0 & 40 & 0\\

24 & 9 & 7.5 & 1 & 7 & 0 & 0 & 0.7 & 0 & 10 & 0 & 0
& $2\times10^3$ & 0 & 40 & 200 & 0 & 40 & 0\\

28 & 9 & 7.5 & 1 & 7 & 0 & 0 & 0.7 & 0 & 10 & 0 & 0
& $2\times10^3$ & 0 & 40 & 200 & 0 & 40 & 0\\

50 & 8 & 8 & 1 & 7 & 0 & 0 & 2 & 0 & 9 & 0 & 0
& $10^4$ & 0 & 10 & 100 & 0 & 20 & 0\\

87 & 6 & 8 & 1 & 7 & 0 & 0 & 10 & 0 & 0 & 0 & 0
& $10^4$ & 0 & 5 & 50 & 0 & 10 & 0\\

115 & 6 & 15 & 1 & 7 & 0 & 0 & 10 & 0 & 0 & 0 & 0
& $10^4$ & 0 & 5 & 50 & 0 & 10 & 0\\

\enddata
\end{deluxetable}


\clearpage

\begin{references}

\reference{} Branch, D., Baron, E., \& Jeffery, D. J. 2003, in
Supernovae and Gamma--Ray Bursters, ed. K.~Weiler (Heidelberg:
Springer), 47

\reference{} Branch, D., Lacy, M. L., McCall, M. L.,
Sutherland,~P.~G., Uomoto,~A., Wheeler,~J.~C., \& Wills,~B.~J. 1983,
ApJ, 270, 123

\reference{} Branch, D., et al. 2003, AJ, 126, 1489

\reference{} Filippenko, A. V. 1997, in Thermonuclear Supernovae,
eds. P.~Ruiz--Lapuente, R.~Canal, \& J.~Isern (Dordrecht: Kluwer), 1

\reference{} Gerardy, C. L., et al. 2004, ApJ, 607, 391

\reference{} Hatano, K., Branch, D., Fisher, A., Baron,~E., \&
Filippenko,~A.~V. 1999a, ApJ, 525, 881

\reference{} Hatano, K., Branch, D., Fisher, A., Millard,~J., \&
Baron,~E. 1999b, ApJS, 121, 233

\reference{} H\"oflich, P. 1995, ApJ, 443, 89

\reference{} Jeffery, D. J. \& Branch, D. 1990, in Supernovae,
eds. J.~C.~Wheeler, T.~Piran, \& S.~Weinberg (Singapore: World
Scientific). 149

\reference{} Jeffery, D. J., Leibundgut, B., Kirshner, R. P.,
Benetti,~S., Branch,~D., \& Sonneborn,~G. 1992, ApJ, 397, 304

\reference{} Kirshner, R. P., et al. 1993, ApJ, 415, 589

\reference{} Lentz, E. J., Baron, E., Branch, D., \&
Hauschildt,~P.~H. 2001, ApJ, 557, 266

\reference{} Mazzali, P. A., et~al. 2005, MNRAS, 357, 200

\reference{} Mazzali, P. A., Lucy, L. B., Danziger, I. J., Gouiffes,
C., Cappellaro, E., \& Turatto, M. 1993, A\&A, 269, 423

\reference{} Meikle, W. P. S., et al. 1996, MNRAS, 281, 263

\reference{} Pastorello, A., et al. 2004, MNRAS, in press

\reference{} Patat, F., Benetti, S., Cappellaro, E., Danziger,~I.~J.,
Della Valle,~M., Mazzali,~P.~A., \& Turatto,~M. 1996, MNRAS, 278, 111

\reference{} Thomas, R. C., Branch, D., Nomoto, K., Li, W., \&
Filippenko,~A.~V. 2004, ApJ, 601, 1019

\reference{} Wang, L., et al. 2003, ApJ, 591, 1110

\end{references}
\end{document}